# The cuttlefish *Sepia officinalis* (Sepiidae, Cephalopoda) constructs cuttlebone from a liquid-crystal precursor

Antonio G. Checa[1,2*], Julyan H.E. Cartwright[2], Isabel Sánchez-Almazo[3], José P. Andrade[4], Francisco Ruiz-Raya[5]

[1]Departamento de Estratigrafía y Paleontología, Universidad de Granada, 18071 Granada, Spain.

[2]Instituto Andaluz de Ciencias de la Tierra, CSIC-Universidad de Granada, 18071 Granada, Spain. [3]Centro de Instrumentación Científica, Universidad de Granada, 18071 Granada, Spain. [4]Centro de Ciências do Mar do Algarve, Universidade do Algarve, Faro, Portugal. [5]Departamento de Zoología, Universidad de Granada, 18071 Granada, Spain. e-mail: acheca@ugr.es

Correspondence and requests for materials should be addressed to A.G.C. (email: acheca@ugr.es).

Cuttlebone, the sophisticated buoyancy device of cuttlefish, is made of extensive superposed chambers that have a complex internal arrangement of calcified pillars and organic membranes. It has not been clear how this structure is assembled. We find that the membranes result from a myriad of minor membranes initially filling the whole chamber, made of nanofibres evenly oriented within each membrane and slightly rotated with respect to those of adjacent membranes, producing a helical arrangement. We propose that the organism secretes a chitin–protein complex, which self-organizes layer-by-layer as a cholesteric liquid crystal, whereas the pillars are made by viscous fingering. The liquid crystallization mechanism permits us to homologize the elements of the cuttlebone with those of other coleoids and with the nacreous septa and the shells of nautiloids. These results challenge our view of this ultra-light natural material possessing desirable mechanical, structural and biological properties, suggesting that two self-organizing physical principles suffice to understand its formation.

$T$he cuttlebone of the European common cuttlefish *Sepia officinalis* Linnaeus, 1758 is an intricate structure composed of a dorsal shield and ventrally placed chamber complex. It is composed of calcium carbonate in its aragonite polymorph mixed with a small amount (3-4.5%[1]) of organic matter, a complex of β-chitin and protein[1-5]. In ventral plan view, the chamber complex consists of the posterior siphuncular zone, which is characterized by a series of striae corresponding each to the posterior end of one chamber, and the septum of the last-formed chamber (Fig. 1a). In dorsoventral vertical section, each chamber is composed of a complex arrangement of horizontal septa and membranes and vertical pillars and membranes (Fig. 1b), which were first described at the end of the 19th century[6], and, in more detail by later authors with the aid of scanning electron microscopy (SEM)[7,8]. All earlier observations[7,8] were based solely on low-resolution SEM, which did not permit a complete recognition of the ultrastructure, and the study of the organic elements was never addressed. The only chamber formation cycle model[8] is therefore far from being fully explicative. For these reasons, there is still much to clarify on both the formation of the different elements of the chamber and their mutual arrangement.

Compared to ectocochleate cephalopods (Nautiloidea and the extinct Ammonoidea), only the septa of the Sepiida can at first glance be homologized, although they are nacreous in the ectocochleates and composed of horizontal aragonite fibres (type-2 nacre[9] or *Spirula*-type nacre[10]) in the Sepioidea. The chamber elements of the Sepiida appear highly modified even when compared with those of the closely related Spirulida[11,12]. The siphuncle of the former is restricted to a series of organic membranes placed posteriorly of the chambers, whereas in the Spirulida it is formed by apicalward tubular extensions of the septa. Pillars are also found in *Spirula* connecting the siphuncular segments of the septa, although it is unclear that they are homologous with those of *Sepia*. In conclusion, the cuttlebone appears unique among cephalopods. An understanding is therefore important, not only with regards to the structure itself, but also to the homologies of its constituent elements with those of other cephalopods and, hence, to the evolution of the cephalopod shell.

We have carried out an in-depth study of the chambers of the common European cuttlefish *Sepia officinalis*, with the aid of high resolution SEM, transmission electron microscopy (TEM) and other analytical techniques. Our results allow us to propose a new model for the formation of the chambers. This new chamber formation model is mediated by a physical process of liquid crystallization, which explains the main features of the *Sepia* chambers and seems to be a common feature in the shells of other cephalopods.

Cuttlebone is a natural material possessing an exceptional combination of desirable mechanical properties of high compressive strength, high porosity and high permeability. These properties are sought after both in biomimetic and in biomedical structural materials. Our findings represent a fundamental step in comprehending the construction of this biogenic material and elucidate how organisms utilize physical principles to generate complex structures[13,14,15].

## Results

**Main elements of the chamber.** The chambers are composed of horizontal septa and membranes, and vertical pillars and membranes. The septa are divided into a chamber roof and a chamber floor. In vertical section, the floor is horizontally layered, whereas the roof is made up of vertical aragonite needles (Fig. 2a). The floor of one chamber lies on top of the roof of the subsequently formed chamber, both being totally contiguous[7] (Fig. 2a).

The pillars extend vertically and are corrugated plates with thicknesses typically of 2-3 µm. Their horizontal cross-sectional outlines change progressively from undulating or labyrinthine at the dorsum of the chamber, where they elongate and tend to align anteroposteriorly (Fig. 2b; Supplementary Video S1), to quasi-dendritic towards the ventral chamber side (Fig. 2c). The surface of the pillars in contact with the chamber floor is sculpted by densely spaced knobs, which tend to be placed at the edges (Fig. 2d). The aragonite needles of the chamber roof are continuous into the tops of the pillars (Fig. 2a), whereas the contact of the pillars with the organic uppermost layer of the chamber floor is loose (Fig. 2e). Pillars show a series of horizontal regularly spaced lines, which have usually been interpreted as growth lines (e.g., [1,7]) (Fig. 2f, g). Observation of polished samples shows that the pillars of successive chambers appear at consistent positions in different chambers (Fig. 2h). Vertical membranes can also be observed extending between pillars (Fig. 2b).

The chamber interior shows a variable number (4 to 10) of horizontal membranes, which extend throughout the chamber, although they tend to split before reaching the pillars (Fig. 1b, 2f, g). The pillars themselves are also surrounded by membranes (peripheral membranes), which seem to be continuous into the horizontal and vertical membranes (Fig. 2g).

**Ultrastructure of the pillars and of the chamber interior.** Slightly calcified (see below) fractured pillars display an internal arrangement of apparently disordered organic fibres (~20 nm thick under SEM) and mineral particles (Fig. 3a). However, when decalcified pillars are observed under TEM, the arrangement is much more ordered. The internal substructure consists of horizontal bands or layers, which are in turn constituted by nanolaminae formed by aligned tiny fibres, the widths of which are in the range of a few nanometers (nanofibres) (Fig. 3b-g). In some cases, the nanolaminae change in orientation, thus producing concentric arches aligned parallel to the layers (Fig. 3d-f). In other cases, the patterns are instead composed of bands with horizontal or inclined laminae, or with a more fuzzy aspect (Fig. 3g).

TEM examination of decalcified samples clearly shows that the individual laminae can be traced from the interior through to the exterior of the pillars (Fig. 3d-e).

Except for one specimen, in which the whole initial set of membranes within the interior of the most recently formed chamber was preserved intact (Fig. 3h), we see that the membranes tend to merge together outside the pillars (Fig. 3b-e), which is clearly a surface-tension effect produced during dehydration of the chamber. For similar reasons, the organic membranes also adhere around the pillars (Fig. 3b, c, e), forming an organic carpet, which we call here peripheral membranes (Fig. 3a). The disordered fibrous aspect of horizontal membranes under SEM (Fig. 3i) is similar to that of fractured pillars (Fig. 3a), although it may be enhanced by the fact that they are multiple, merged membranes.

The membranes internal to the pillars also become convex towards the venter, throughout the length of the pillars (Fig. 3j, k). The curvature extends to the portions of the membranes outside and immediately adjacent to the pillars (Fig. 3l).

**The septum.** Vertical fractures reveal that the chamber floor is composed of thin aragonite fibres (100-200 nm thick), which are distributed within planes (Fig. 4a). Within each plane, the fibres are evenly oriented, but they have different orientations (at high angles) in the different planes (Fig. 4a, b).

The chamber floor is covered by a dorsal layer of nanofibres, which is not calcified. The calcified septum is found below this fibre layer (Fig. 4c). TEM images of decalcified specimens show that the chamber floor is formed by particularly thick organic fibres, which may exceed 100 nm in diameter (Fig. 4d-f). Horizontal sections of the septum reveal that, in a way similar to its aragonite fibres, the organic fibres are also distributed in a series of planes; within each plane the fibres are parallel to each other, but rotated at a relatively high angle with respect to the fibres in the adjacent planes (Fig. 4d, e). In vertical sections the fibres are distributed into bands, which show a typical arcuate distribution of fibres (Fig. 4f). TEM of non-decalcified specimens shows that the aragonite fibres are identically patterned (Fig. 4c), which implies that the aragonite and organic fibres run in parallel. In one juvenile specimen we counted 15-18 such bands, each having a thickness between 200-400 nm. In non-calcified (see below) or decalcified specimens, the thick organic fibres of the floor layer change into the much more densely distributed and thinner horizontal nanofibres of the roof layer (Fig. 4e).

In detail, the septum has a smoothly undulated upper surface, becoming concave towards the preceding chamber at the positions of the pillar edges and slightly convex at the interspaces (Figs. 3j-l). Above the pillar edges, the septum thickness becomes reduced to almost half the usual value, although the number of layers remains the same. The undulations of the septum fit in with those of the horizontal membranes trapped within the pillars, as described above.

**The siphuncular area.** At the siphuncular zone, the pillars become progressively shorter and the whole arrangement of membranes converges towards the siphuncular end of the chamber, according to the wedge-like shape of the chamber profile (Fig. 4g). The pillars lose their dendritic outlines and become columnar to irregularly elongate in the anteroposterior direction (Fig. 4h). They sometimes divide and acquire an antler-like morphology (Supplementary Video S1). As in the rest of the chamber, the organic bands run uninterruptedly across the pillars, but with the peculiarity that they are totally flat and do not undulate at the positions of the pillars (Fig. 4i). The siphuncular septum is merely a particularly dense accumulation of membranes, which (contrary to those of the non-siphuncular septum) are identical in diameter and patterns to those of the chamber interior (Fig. 4i). The increase in density of organic bands is not abrupt and it is difficult to set exactly the boundary between the siphuncular septum and the chamber interior (Fig. 4i).

**Non-calcified chambers.** In juvenile specimens the degree of calcification of the pillars and the rest of the chamber structures becomes progressively more incipient in the last 3-6 chambers. Different specimens show different patterns of calcification. In the instance shown in Supplementary Fig. S1, the pillars and septa are totally uncalcified in the last chamber; in the penultimate chamber the chamber roof is already calcified while the pillars only show calcified patches, which are more extensive towards their dorsal parts. All elements, including the chamber floor, are fully calcified in the antepenultimate chamber. Progressive calcification is also demonstrated with Raman spectroscopy (Supplementary Fig. S1). In other cases, the calcification is not yet complete in the last five chambers. The last-formed chambers of some relatively aged specimens were apparently calcified, although it is not clear that they were actively growing at the time of capture. In summary, pillars and septa can be observed in all states of calcification in the last formed chambers: purely organic (Fig. 5a), very incipiently (Fig.

5b), slightly (Fig. 5c), moderately (Fig. 5d) or fully calcified (Fig. 2e, f). The same trends are revealed by TEM. Observation of an apparently uncalcified pillar (similar to that shown in Fig. 5a) shows the existence of very sparse calcium carbonate granules. With increasing degrees of calcification (Fig. 5f), the density of such particles also increases. They typically align with the organic membranes and are 50-150 nm in size (Fig. 5g, h). Electron diffraction demonstrates that they are fully crystalline (Fig. 5h). With advancing calcification, the density and size of particles increases (Fig. 5i). During the calcification process, the pillars are always found in a more advanced state of calcification than the septa (Fig. 5i).

## Discussion

Our new results can be summarized as follows: (a) The chambers are filled with a sequence of layers or bands, made of fibrous nanolaminae, which display a horizontal or arcuate patterning. (b) These layers run uninterruptedly across the pillars. At their positions, the layers become convex in a ventral direction. (c) The chamber floor part of the septum is composed of an arrangement of organic fibres, which are identical to, although much thicker than, those composing the chamber interior and the chamber roof. (d) The uncalcified siphuncular part of the septum is simply a particularly dense accumulation of membranes identical to those of the chamber interior. (e) Pillars and septa begin with a purely organic stage, and calcification proceeds afterwards. Pillars and the chamber roof part of the septum are continuously mineralized with aragonite fibres oriented parallel to the pillar axis, whereas in the chamber floor the aragonite fibres follow the orientation of the organic fibres. These observations have clear implications with regards to the ultrastructure and mode of formation of chambers.

The organic matter present within the cuttlebone is a mixture of chitin and protein[1,2] in the form of complexes[3,16]. The percent of chitin to protein within the cuttlebone was estimated to be about 50%[3] and 30%[5]. Chitin has been determined to be is in its β-polymorph[4,5], which is ubiquitous in molluscs[17]. In our samples, the arcuate patterns observed are formed by electron-lucent (i.e., non- to poorly stained) nanofibres (chamber interior and roof) or microfibres (chamber floor). Of the two components, only the protein has the ability to stain. Therefore, and judging also from the case of arthropod cuticle (discussed in, e.g.,[18]), the pattern reflects the distribution of the chitin units. The 2D arched aspect of fibres has been found extensively in plants, arthropods and other groups[19]. The arcuate patterns seen in the crustacean cuticle have been convincingly interpreted as the oblique sectional view of a 3D arrangement of fibres distributed in superposed planes, with fibres within each plane being parallel and rotated by a small angle with respect to those in the adjacent planes[20]. The helical structure of insect cuticles has since long been assimilated to the cholesteric phase of a liquid crystal[21-24] made, in that instance, of the $\alpha$-polymorph of chitin. Liquid-crystal self-organization has been postulated to explain the structural plywood arrangement of many other fibrous biocomposites, from plant cell walls to fish scales and bone[19,22]. Such an arrangement has been found to have an impressive biomechanical performance (e.g.,[25]). Helical arrangements have been described in the protein-chitin complex forming the pen of the squid *Loligo*[26,27], which is homologous to the dorsal shield of the cuttlebone. The same liquid-crystal organizing principle was employed to explain the formation of the interlamellar membranes of molluscan nacre, although there the constituent fibres within each membrane are disoriented[28,29].

The increasingly dendritic shapes of the pillar edges are particularly appealing. An interpretation in which the progressive complication of the margin of the organic pillars, which is reproduced periodically during the formation of each chamber, results from a genetically-based complex behaviour of particular groups of cells is possible, but hardly understandable. A much simpler alternative is to explain pillar shape development by means of the so-called viscous fingering instability[30]. This happens at the interface between two fluids of different viscosities when the pressure of the less viscous fluid exceeds that of the more viscous fluid.

The resulting interface then becomes dendritic, thus forming the so-called viscous fingers (Fig. 6a, b). The importance of viscous-fingering formation in developmental biology and fabricational morphology has been stressed[31,32]. In fact, within the forming cuttlebone chambers, there are two intervening fluids: the (lower viscosity) cameral liquid and the (more viscous) mucous (possibly proteinaceous) precursor of the pillars that we have demonstrated to exist prior to calcification. To complete the biophysical scenario we need evidence that the pressure within the organic pillars could have been less than that of the cameral fluid. As described above, the chamber-filling membranes and the septum floor are relatively depressed (convex ventralwards; Fig. 3j-l) at the positions of the pillars, which implies that the mantle was also depressed at the same positions during the secretions of the organic precursors of the pillars. After the secretion of the septum, the mantle surface largely, though not fully, levels off at the pillar and inter-pillar positions, as deduced from the shape of the bottom surface of the septum roof (Fig. 3l). In other words, during secretion of the pillars, their edges are pulled by the mantle, thus causing progresive deformation of the horizontal membranes. Pulling of the pillar precursor by the soft body would bring about a certain drop of its internal pressure, which would then be intruded by the cameral liquid. Accordingly, the pillar precursor would begin to develop viscous fingers at its edge, and the pillar surface would acquire a general anticlastic morphology. Some labyrinthine patterns produced by pillars close to their dorsal ends can be successfully reproduced experimentally (Fig. 6a-c). Since at the dorsal edge, the organic pillars are continuous with the organic precursor of the chamber roof, which extensively carpets that surface, the situation we encounter is very similar to the experimental one (see caption of Fig. 6 for experimental details). The viscous fingering instability that we are concerned with here is the so-called lifting version, found when a less viscous fluid displaces a more viscous one between two separating plates[33]. We cannot at present be certain as to the origin of these differences observed in the viscous fingering pattern at the roof and the floor of each chamber, where the outlines of pillars (which we have failed in reproducing; Figs. 2c, 6d) are noticeably dendritic. However, as we have discussed, the roof and floor have different structures and compositions. We speculate, then, that the different morphologies reflect different amounts of `stickiness' of the two surfaces. In classical studies of viscous fingering the interface is supposed to slide smoothly under pressure; if, as we suspect to be the case here, there are differing amounts of interface pinning at the top and bottom boundaries, this would account for the strikingly different morphologies observed at the bases and the tops of the pillars. The fact that pillars are found in exactly the same positions in successive chambers[6] (Fig. 2h) argues for the existence of particular areas of the mantle epithelium specialized in the secretion of such a proteinaceous precursor.

Since we have observed that membranes go uninterruptedly across the pillars (e.g. Fig. 3b-e), it is hardly understandable that chitin fibres can arrange themselves homogeneously across both the low (cameral fluid) and high (mucous precursor of the pillars) viscosity fluids. This objection is overcome if, instead of a suspension of disorganized fibres filling the chamber, which organizes with time, we think of a liquid crystal growing layer by layer (as proposed for the interlamellar membranes of nacre[28]). In this way, the cholesteric liquid crystal does not organize at once but progressively as the chamber is constructed.

Although all the details of the calcification process are not known, it begins with nanometric crystalline particles of calcium carbonate, which arrange onto the organic matrix formed by parallel nanomembranes (Fig. 5g). Both the chamber roof and pillars are calcified with aragonite needles, which run perpendicular to the floor and parallel to the axes of the pillars (Fig. 2a, e). In the floor part of the septum, on the contrary, the aragonite fibres run horizontally and change their directions in the different planes (Fig. 4a, b, c) according to the orientation of the chitin fibres (Fig. 4d-g). In the roof and pillars, the orientation of the aragonite fibres can be explained by the electrostatic interactions between the positively charged calcium planes of aragonite, perpendicular to the crystallographic $c$-axes (i.e., the elongation axes of needles), and the negatively charged (horizontal) protein laminae. Similar processes are common in biomineralization (see review in[34]). The orientation of the aragonite fibres of the septum floor in parallel to the thick chitin fibres is not at present understood, unless there is an intermediate protein bound to the chitin, which might specifically orient the aragonite fibres (e.g.,[35]). This particular orientation of the septal fibres, leading to a typical plywood structure, is probably a mechanical adaptation to provide a better performance against hydrostatic load.

According to our new evidence, the chamber formation cycle would consist of the following stages: (1) Secretion of the septum roof, which would initially consist of slightly undulating chitin/protein laminae plus the organic proteinaceous precursor of this layer. Secretion is later restricted to laminae, with the organic precursor mineralizing substance only being present at the positions of the pillars, which are, in this way, continuous into the chamber roof (Fig. 7a). Vertical membranes subtending between pillars are also secreted at this stage. Throughout this process, the fibres of each successive nanolamina become immediately ordered onto those of the previous one following a cholesteric liquid-crystal pattern. The ventral ends of the forming organic pillars are pulled by the mantle, which thus bends ventralwards at these sites. The resulting viscous-fingering formation process causes the edges of the pillars to acquire their typical dendritic shapes. This process does not take place at the siphuncular zone. Accordingly, the siphonal pillars retain club-like shapes (Fig. 4g, i). (2) Septum secretion follows (Fig. 7b). This is composed of an arrangement of thick chitin fibres also forming a cholesteric phase with a pseudo-orthogonal arrangement[19] (Fig. 4d-f). After secretion of the septum the high curvature zones formerly developed b y the mantle onto the pillar surfaces revert to smoother. Contrary to the septum, the siphuncular membrane is made of densely space membranes composed of nanofibres of the type found within the chamber interior (Fig. 4i). (3) Mineralization of the chamber proceeds with the formation of crystalline granules, first within the chamber roof to later extend to the pillars until they are completely mineralized (Fig. 7c). The last structure to be mineralized is the septum floor, where the aragonite fibres, unlike those of the septum roof and pillars (vertical), are arranged horizontally and in parallel to the chitin fibres of the superposed planes. The siphuncular membrane does not mineralize at all. (4) Drainage of the chamber, which causes the formation of secondary membranes resulting from the merging of the original nanomembranes (Fig. 7d).

The cuttlebone constitutes a highly efficient buoyancy apparatus, evolved to withstand hydrostatic pressure[36]. Our study reveals that in *Sepia* the construction of chambers relies on physical processes of liquid crystallization and viscous fingering. On this basis, it is possible to understand how such a complex structure can self-organize making use of relatively simple physical principles.

Liquid crystallization seems to be ubiquitous within coleoid cephalopods. It has been invoked to explain the similarly arcuate patterns observed also in the pen of the squid *Loligo*[26,27]. We have observed similar patterns in the dorsal shield of the cuttlebone (which is homologous to the squid pen) and in the septa of *Spirula* (Supplementary Fig. S2). Accordingly, shell and septa construction by liquid crystallization has been conserved at least prior to the divergence of squids (Myopsida) from the Sepiida/Spirulida (all these groups being grouped under the Decapodiformes), dating at least ca. 117 mya (when the oldest sepioid is recorded[37]), although fossil teuthoids could date back to the Upper Triassic[14,38] (~215 mya). Liquid crystallization has also been invoked to explain the formation of the interlamellar membranes of molluscan nacre[28]. Therefore, this constructional strategy can be traced even further back to at least the Upper Ordovician, when the first nautiloid nacre is recorded[39,40] (~445 Ma). Accordingly, we may conclude that the nacreous septa of nautiloids are homologous to the septa of decapodiforms, and the partly nacreous shells of nautiloids are
homologous to the squid pen and to the dorsal shield of the cuttlebone.

## Methods

**Ethics Statement.** Rearing and handling of specimens were performed in accordance with
relevant guidelines and regulations, under certification to the Centro de Ciências do Mar
(University of Algarve) issued by the Direcção Geral de Veterinaria (Ministério da Agricultura,
do Desenvolvimento Rural e das Pescas, Governo de Portugal). J.P.A. is also certificated for the
same activity by the latter Institution.

**Materials.** Very young specimens of *Sepia officinalis* (up to fourteen days after hatching)
reared at the Centre of Marine Sciences, University of the Algarve (Faro, Portugal), were fixed
*in vivo* with 2.5% glutaraldehyde buffered with cacodylate (0.1 M, pH 7.4). Other specimens
with commercial sizes were purchased at local fish markets in Granada. The dissected
cuttlebones, together with their secreting ventral epithelium, were fixed in the same solution.
Other, dry cuttlebones were also used for complementary observations. A series of samples
was also wholly decalcified by immersion in either 2% EDTA.

**Optical, electron microscopy and nano-CT.** Prior to observation, pieces of the cuttlebone and
the accompanying ventral epithelium of the gluta-fixed specimens were $CO_2$-critical-point
dried (Polaron CPD 7501). Specimens observed in SEM were previously coated with carbon
(Hitachi UHS evaporator). We used the field emission SEMs Zeiss LEO Gemini 1530 and Zeiss
Auriga Cross-Beam Station and the Environmental SEM (ESEM) FEI Quanta 400 of the
University of Granada. Specimens for optical microscopy (OM) and TEM (some of which were
completely decalcified by immersion in 2-4% EDTA) were post-fixed in $OsO_4$ (2%) for 2 h at 4 °C
and embedded in epoxy resin Epon 812 (EMS). They were sectioned with an ultramicrotome
LEICA Ultracut R and prepared following standard procedures. Semi-thin sections (~0.5 μm)

were stained with 1% toluidine blue and observed with an Olympus BX51 microscope. Ultra thin sections (50 nm) were stained with uranyl acetate (1%) followed by lead citrate. They were later carbon-coated and observed with TEM (Zeiss LEO 906E, Zeiss Libra 120 Plus and FEI Titan at the University of Granada). The TEM Zeiss Libra Plus 120 is equipped with an in-column Omega filter, which is able to carry out elemental analysis by means of Electron Energy Loss Spectroscopy (EELS) and produce detailed element distribution maps (Energy Spectroscopic Imaging, ESI).

The last formed chambers of young cuttlefishes were cut into very small pieces for 3D reconstruction. Scanning was made with a Skyscan micro-CT attachment installed on an ESEM FEI Quanta 400, University of Granada). X-rays for tomographic analysis were generated with an electron beam acceleration voltage of 30 kV. The rotation steps were 0.45° up to a total rotation of 180°. The pixel size is 1.1mm. The Bruker Skyscan free software ®NRecon, ®CTNn, ®DataViewer and ®CTvox were used to reconstruct and process the images, and to generate the video.

**Raman spectroscopy.** Raman spectra were recorded using a Jasco NRS-5100 Spectrometer equipped with a 4-stage Peltier cooled CCD detector. The laser radiation was blocked using edge filters and the scattered light was dispersed by a grating with 1800 grooves/mm. Microscopic observation was made with an integrated high-resolution built-in CMOS camera (100x  magnification) with automated stage. A laser wavelength of 532.04 nm was used for excitation with a laser power of 50 mW. Raman spectra were recorded with 5 s acquisition times after five accumulations. Laser spot size was 1 μm.

# References

1.  Birchall, J. D. & Thomas, N. L. On the architecture and function of cuttlefish bone. *J. Mater. Sci.* **18**, 2081-2086 (1983).

2.  Hare, P. E. & Abelson, P. H.. Amino acid composition of some calcified proteins. *Carnegie Inst. Washington Year Book* **64**, 223-232 (1965).

3.  Hackman, R. H. Studies on chitin. IV. The occurrence of complexes in which chitin and protein are covalently linked. *Australian J. Biol. Sci.* **13**, 568-577 (1960).

4.  Hackman, R. H., Goldberg M. Studies on chitin. VI. The nature of α- and β-chitins. *Australian J. Biol. Sci.* **18**, 935-946 (1965).

5.  Okafor, N. Isolation of chitin from the shell of the cuttlefish, *Sepia officinalis* L. *Biochim. Biophys. Acta* **101**, 193-200 (1965).

6.  Appellöff, A. Die Schalen von Sepia, Spirula und Nautilus. Studien über den Bau und das Wachstum. *Kongl. Svensk. Vetensk. Akad. Handl.* **25**, 1-106 (1983).

7.  Bandel, K. & Boletzky, S. v. A comparative study of the structure, development and morphological relationships of chambered cephalopod shells. *Veliger* **21**, 313-354 (1979).

8.  Tanabe, K., Fukuda, Y. & Ohtsuka, Y. New chamber formation in the cuttlefish *Sepia sculenta*. *Venus* **44**, 55-67 (1985).

9.  Mutvei, H. Ultrastructure of the mineral and organic components of molluscan nacreous layer. *Biomineralisation* **2**, 48-61 (1970).

10. Dauphin, Y. Microstructures et flottabilité chez la spirule (Cèphalopoda). *C. R. Acad. Sci. Paris D* **284**, 2483-2485 (1977).

11. Young, R. E. & Vecchione, M. Analysis of morphology to determine primary sister-taxon relationships within coleoid cephalopods. *Amer. Malacol. Bull.* **12**, 91-112 (1996).

12. Young, R. E., Vecchione, M. & Donovan, D. T. The evolution of coleoid cephalopods and their present biodiversity and ecology. *South African J. Marine Sci.* **20**, 393-420 (1998).

13. Wegst , U. G. K., Bai, H., Saiz, E., Tomsia, A. P. & Ritchie, R. O. Bioinspired structural materials. *Nature Mater.* **14**, 23-26 (2015).

14. Sanchez, C., Arrivart, H. & Giraud Guille, M. M. Biomimetism and bioinspiration as tools for the design of innovative materials and systems. *Nature Mater.* **4**, 277-288 (2005).

15.- Cartwright, J. H. E. & Mackay, A. L. Beyond crystals: the dialectic of materials and information. *Phil. Trans. R. Soc. A* **370**, 2807-2822 (2012).

16. Blackwell, J. & Weih, M.A. Structure of chitin-protein complexes: ovipositor of the ichneumon fly *Megarhyssa*. *J. Molec. Biol.* **137**, 49-60 (1980).

17. Weiner, S. & Traub, W. X-ray diffraction study of the insoluble organic matrix of mollusk shells. *FEBS Lett.* **111**, 311-316 (1980).

18. Hackman, R. H. in *Chitin and Benzoylphenyl Ureas* (eds Wright, J. E. & Retnakaran, A.) 1-32 (W. Junk, Dordrecht, The Netherlands, 1987).

19. Neville, A. C. *Biology of Fibrous Composites, Development beyond the Cell Membrane* (Cambridge Univ. Press, Cambridge, UK, 1993).

20. Bouligand, Y. Sur une architecture torsadée répandue dans de nombreuses cuticules d'Arthropodes. *C. R. Acad. Sci. Paris D* **261**, 3665-3668 (1965).

21. Bouligand, Y. Sur l'existence de 'pseudomorphes cholestériques' chez divers organismes vivants. *J. Phys.* **30 Suppl. C4**, 90-103 (1969).

22. Bouligand, Y. Twisted fibrous arrangements in biological materials and cholesteric mesophases. *Tissue & Cell* **4**, 189-217 (1972).

23. Neville, A. C. & Luke, B. M. A two-system model for chitin-protein complexes in insect cuticles. *Tissue & Cell* **1**, 689-707 (1969).

24. Neville, A. C. & Luke, B. M. A biological system producing a self-assembling cholesteric protein liquid crystal. *J. Cell Sci.* **8**, 93-109 (1971).

25. Zimmermann, E. A. *et al*. Mechanical adaptability of the Bouligand-type structure in natural dermal armour. *Nature Comm.* **4**, 2634 (2013).

26. Hunt, S. & El Sherief, A. A periodic structure in the 'pen' chitin of the squid *Loligo vulgaris*. *Tissue & Cell* **22**, 191-197 (1990).

27. Levi-Kalisman, Y., Falini, G., Addadi, L. & Weiner, S. Structure of the nacreous organic matrix of a bivalve mollusk shell examined in the hydrated state using cryo-TEM. *J. Struct. Biol.* **135**, 8–17 (2001).

28. Cartwright, J. H. E. & Checa, A. G. The dynamics of nacre self-assembly. *J. R. Soc. Interface* **4**, 491-504 (2007).

29. Cartwright, J. H. E., Checa, A. G., Escribano, B. & Sainz-Díaz, C. I. Spiral and target patterns in bivalve nacre manifest a natural excitable medium from layer growth of a biological liquid crystal. *Proc. Natl. Acad. Sci. U.S.A.* **106**, 10499-10504 (2009).

30. Saffman, P. G. & Taylor, G. The penetration of a fluid into a porous medium or Hele-Shaw cell containing a more viscous liquid. *Proc. R. Soc. A* **245**, 312-329 (1958).

31. Cartwright, J. H. E., Piro, O. & Tuval, I. Fluid dynamics in developmental biology: Moving fluids that shape ontogeny. *HFSP J.* **3**, 77-93 (2009).

32. Checa, A. G. & García-Ruiz, J. M. in *Ammonoid Paleobiology* (eds Landman, N. H., Tanabe, K. & Davis, R. A.) 253-296 (Plenum Press, New York, NY, 1996).

33. Nase, J., Derks, D. & Lindner, A. Dynamic evolution of fingering patterns in a lifted Hele–Shaw cell. *Phys. Fluids* **23**, 123101 (2011).

34. Weiner, S. & Addadi, L. Design strategies in mineralized biological materials. *J. Mater. Chem*. **7**, 689-702 (1997).

35. Falini, G., Weiner, S. & Addadi, L. Chitin-silk fibroin interactions: relevance to calcium carbonate formation in invertebrates. *Calcif. Tissue Int.* **72**, 548-554 (2003).

36. Denton, E. J. & Gilpin-Brown, J. B. The buoyancy of the cuttlefish *Sepia officinalis* (L.). *J. Mar. biol. Ass. U.K.* **41**, 319-342 (1961).

37. Doguzhaeva, L. A. Two early Cretaceous spirulid coleoids of the north-western Caucasus: their shell ultrastructure and evolutionary implications. *Palaeontology* **39**, 681-707 (1996).

38. Reitner, J. Ein Teuthiden-Rest aus dem Obernor (Kössener-Schichten) der Lahnewies-Neidemachmulde bei Garmisch-Partenkirchen (Bayern). *Paläontol. Z.* **52**, 205-212 (1978).

39. Mutvei, H. Flexible nacre in the nautiloid *Isorthoceras*, with remarks on the evolution of cephalopod nacre. *Lethaia* **16**, 233–240 (1983).

40. Vendrasco, M. J., Checa, A., Heimbrock, W. P. & Baumann, S. D. J. Nacre in molluscs from the Ordovician of the Midwestern United States. *Geosciences* **3**, 1-29 (2013).

## Acknowledgements

Special thanks are given to António Sykes (Univ. Algarve) for his help with breeding cuttlefish. This research was funded by projects CGL2010-20748-CO2-01 (to A.G.C., F.R.R. and I.S.A.), CGL2013-48247-P (to A.G.C.) and FIS2013-48444-C2-2-P (to J.H.E.C.) of the Spanish Ministerio de Ciencia e Innovación, and RNM6433 (to A.G.C., J.H.E.C. and I.S.A.), of the Andalusian Consejería de Innovación Ciencia y Tecnología, and 31.03.05.FEP.002 (Sepiatech, PROMAR program) of the Portuguese Ministério da Agricultura e do Mar, Portugal (to J.P.A.). A.G.C. and J.H.E.C. also acknowledge the Research Group RNM363 (Junta de Andalucía) and the FP7 COST Action TD0903 of the European Community.

## Author contributions

A.G.C. conceived, designed and coordinated the study, participated in data acquisition and analysis and drafted the manuscript; J.H.E.C. participated in the design of the study and revised it critically for important intellectual content; I.S.A. participated in data acquisition and analysis; J.P.A. provided essential material and participated in data acquisition; F.J.R.R. participated in the design of the study and in data acquisition and analysis. All authors checked the draft manuscript and gave final approval for publication.

## Additional information

**Supplementary information** accompanies this paper at

http://www.nature.com/scientificreports

**Competing financial interests:** The authors declare no competing financial interests.

**Figure legends**

**Figure 1 | Structure of cuttlebone. a**, Ventral view of the cuttlebone. **b**, Lateral view of a chamber showing the main constituting elements. ds= dorsal shield, hm= horizontal membrane, ls= last septum, p= pillar, s= septum, s= septum, sz= siphuncular zone, vm= vertical membrane.

**Figure 2 | General structure of the chambers. a**, Fracture view of the septum and the initiation of one pillar. The chamber floor is the lower layered half and the roof is the upper half made of vertical aragonite needles and continuous into the pillar. **b**, View of a juvenile chamber from the venter showing anteroposteriorly aligned pillars and adjoining vertical membranes. **c**, View of the ventral surfaces of the pillars, showing their dendritic outlines. **d**, Detail of the ventral surface of one pillar showing the decoration consisting of knobs placed at the edges. **e**, Fracture through the ventral part of a pillar and the septum, showing the loose attachment of both elements, as well as the aragonite fibres composing the pillar. **f**, Lateral view of the ventral side of a pillar and its adjacent septum. Note horizontal partitions and horizontal organic membranes. **g**, Broken ventral end of a pillar showing the organic envelope (peripheral membrane), which is continuous into the horizontal membranes. **h**, Section through an embedded and polished juvenile specimen. Note general correspondence between the positions of the pillars of the different chambers. The venter is to the top in **a**, **e**, **f**, **g** and **h**. All are secondary electron SEM images, except for **f** (backscatter SEM image). chf= chamber floor, chr= chamber roof, ds= dorsal shield, hm= horizontal membrane, p= pillar, pm= peripheral membrane, s= septum, vm= vertical membrane.

**Figure 3 | Structure of cameral elements. a**, Fracture of the vener of an incipiently calcified pillar revealing its internal ultrastructure made of organic threads and mineral (calcium carbonate) particles. **b**, **c**, Sections through the ventral areas of decalcified pillars, showing the existence of internal membranes. These extend outwards from the pillars, where they merge and form horizontal (**b**, **c**) and peripheral (**c**) membranes. **d**, **e**, Aspect of two decalcified pillars close to the siphonal area, in which the distribution of layers can be discerned according to changes in contrast. Each layer is made of nanofibres, which are distributed in arches and can be followed from the interior to the exterior of the pillar (where they tend to merge forming horizontal membranes). **f**, Detail of the layers seen in **d**. **g**, Set of laminae outside the pillars. They are discernible after the orientation of their constituent fibres; an arcuate pattern is not evident. **h**. Section of the last formed chamber of a juvenile specimen. The original distribution of laminae has been preserved, without any merging effect. **i**, Detail of the fibrous aspect of a horizontal membrane. **j-l**, Vertical sections through the penultimate septum and its adjacent pillars in a decalcified sample. Note the changes in thickness of the septum at the contacts with the pillars of the penultimate chamber (bottom parts of the photographs). Where the laminae internal to the pillars are evident, they are always concave dorsalwards (i.e. towards the bottom of the photographs). **l** is a detail of the contact of the pillar with the septum, where the individual nanolaminae are particularly evident. The venter is to the top in **b**, **d**, **e**, **g**, **h** and **k**, to the top left in **f**, **j** and **l**, and to the right in **c**. **a** and **i** are secondary electron SEM image and **h** is a backscatter SEM image; **b** to **g**, **j** and **l** are TEM images; **k** is an optical microscopy image. chf= chamber floor, chr= chamber roof, hm= horizontal membrane, l= laminae, mp= mineral particle, of= organic fibres, p= pillar, pm= peripheral membrane, s= septum.

**Figure 4 | Structure of cameral elements. a,** Oblique fracture of the chamber floor part of the septum showing the distribution of aragonite fibres into planes. Fibres within a plane have

even orientations, and different from those of adjacent planes. **b**, Surface view of a horizontally fractured chamber floor showing two planes of aragonite fibres at a high angle. **c**, Transition from the chamber floor to the chamber interior in a dorsalward direction (bottom right). The transition is marked by a drastic reduction of the sizes of the fibres. **d**, Horizontal section through the septum of a decalcified sample. Owing to the undulated morphology of the septum, the section contains the pillar, and the chamber floor and roof. **e**, Detail of **d**, showing the transition from the chamber floor to the chamber roof. It is marked by a sudden decrease of the diameters of the fibres. **f**, Vertical section through the decalcified chamber floor, showing the arcuate arrangement of fibres. **g**, Vertical fracture through the siphuncular part of the chamber. The septum of the preceding chamber is visible at the bottom. **h**, Ventral view of the siphuncular part of the chamber, showing the irregularity of the pillars in this area. **i**, Section through the pillars and siphuncular membrane of a decalcified sample. The latter is merely a dense accumulation of horizontal membranes. The venter is to the top in **g** and **i**, to the top left in **c**, and to the top right in **f**. **a**, **b**, **g** and **h** are backscatter SEM images; **c** to **f** and **i** are TEM images. chf= chamber floor, chr= chamber roof, fl= fibre layer, hm= horizontal membrane, p= pillar, s= septum, sps= siphuncular septum.

**Figure 5 | Pillars and septa in different states of calcification. a**, **b**, Fully organic pillars (**a**) and very incipiently calcified (white bands) pillars (**b**) and associated vertical membranes found in the last-formed chambers of juvenile specimens. **c**, Slightly calcified pillars found in the penultimate chamber of a juvenile specimen. Note their flexible behaviour. **d**, Pillar in an advanced, though not final state of calcification. Note its fibrous aspect. **e**, Section through an uncalcified pillar of the penultimate chamber of a juvenile specimen. TEM reveals the existence of very sparsely distributed calcium carbonate particles (arrows). The inset is a close up of one such particle. **f**, Section through a slightly calcified pillar showing the alignment of

aragonite particles and organic membranes. The sample has torn during ultramicrotomy (white holes). **g,** EELS calcium map of **f**, showing the distribution of mineral (red) particles. **h,** Detail of particles in and incipiently calcified pillar, similar to that in **f**. The diffraction pattern (inset) implies that particles are fully crystalline. **i,** Section through a pillar in an advanced state of calcification and the semicalcified adjacent septum. The venter is to the top in **e**, to the top left in **i,** and to the bottom right in **f** and **g**. **a** to **d** are backscattered SEM images; **e** to **i** are TEM images. chf= chamber floor, fl= fibre layer, p= pillar, s= septum, vm= vertical membrane.

**Figure 6 | Comparison of adhesion figures to pillars**. **a**, **b**, Two consecutive stages of the formation of an adhesion figure. It has been obtained by placing and pressing an elongated volume of a viscous gel (blue) between two metacrylate plates, which are then pulled apart. Upon pulling, the air intrudes the gel and the interface breaks into a series of viscous fingers. From **a** to **b** the gel has relaxed slightly and the dendrites have shortened and simplified. **c**, Ventral view of the roof of a chamber in which the pillars have become fractured close to the roof. The labyrinthic pattern is pretty close to the experimental viscous fingers, including the existence of bifurcated branches (arrows in **a** and **c**). **d**, Aspect of the ventral edge of the pillars of the preceding chamber, showing the anticlastic morphology of the plates. This morphology cannot be reproduced with our experimental setup, probably due to the strong adhesion of the gel to the plates, which prevents the necessary horizontal displacement of the gel. Field of view for **a** and **b** is *ca.* 12 cm.

**Figure 7 | Proposed cycle for the formation of chambers in *Sepia*. a**, Growth of the chamber. The dorsal epithelium of the mantle secretes a series of organic laminae organized into layers, according to a liquid crystal cholesteric phase, at the same time as the organic precursors of

both the chamber roof and pillars (blue colours). The liquid crystal is secreted layer by layer and the laminae go across the organic pillars, forming horizontal partitions at the intersection (right boxes). **b**, Completion of the chamber. This is achieved with the complete growth of the chamber in height and with the secretion of the organic precursor of the chamber floor. The pillars change in outline from labyrinthine to dendritic in a ventral direction (left boxes). During growth of the chamber, the pillars are progressively stretched by the mantle, which causes an accentuation of both its undulated surface and those of the secreted layers. The mantle reverts to a smoothly undulated shape after secretion of the septum (right boxes), which causes the observed reduction in thickness of the chamber roof at the ventral ends of the pillars. **c**, Initiation of a new chamber cycle and onset of calcification in the previous chamber. This proceeds with the formation of calcium carbonate particles onto the organic laminae of the pillars (right boxes) and septa. **d**, Complete calcification and subsequent drainage of earlier chambers. Upon desiccation, primary organic layers and laminae merge to form secondary horizontal and peripheral membranes (right boxes).

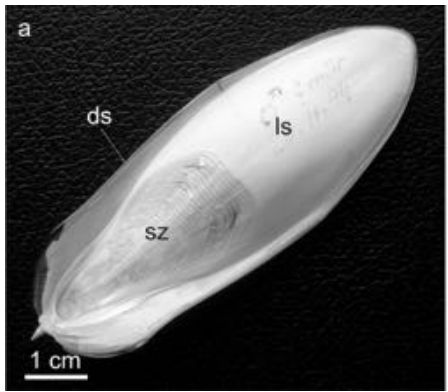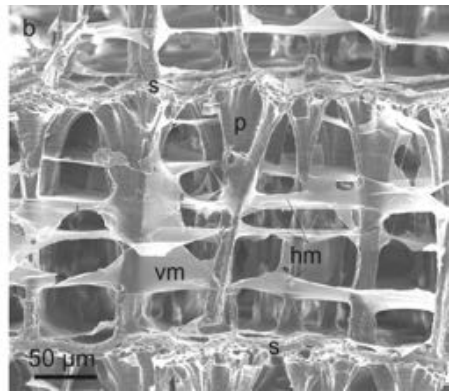

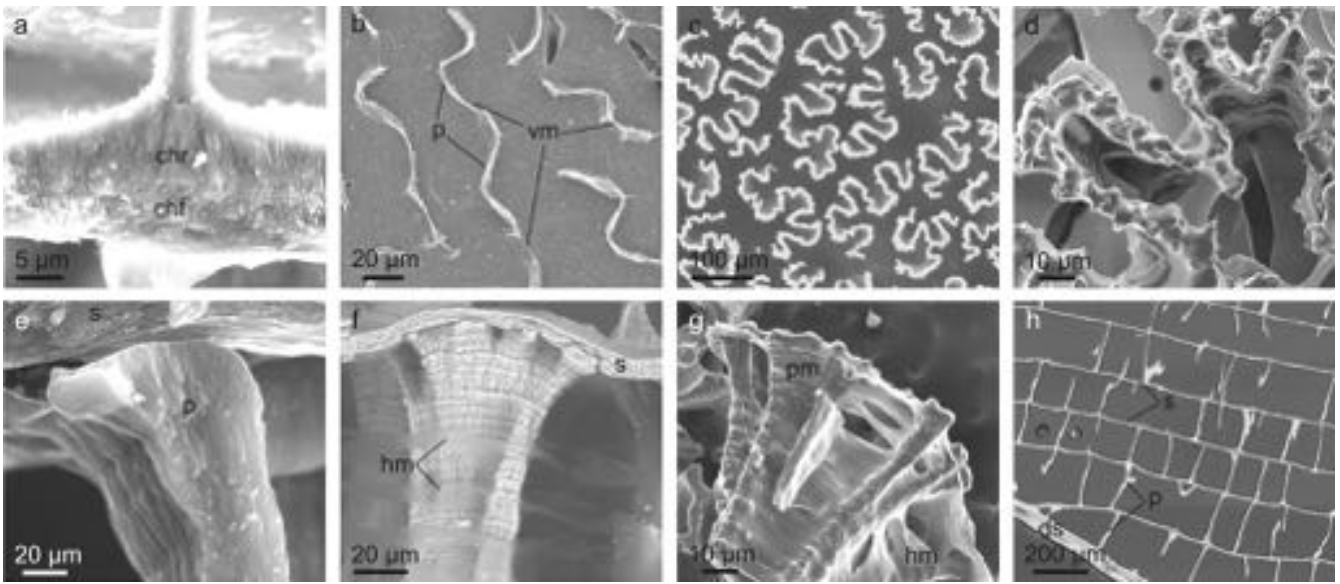

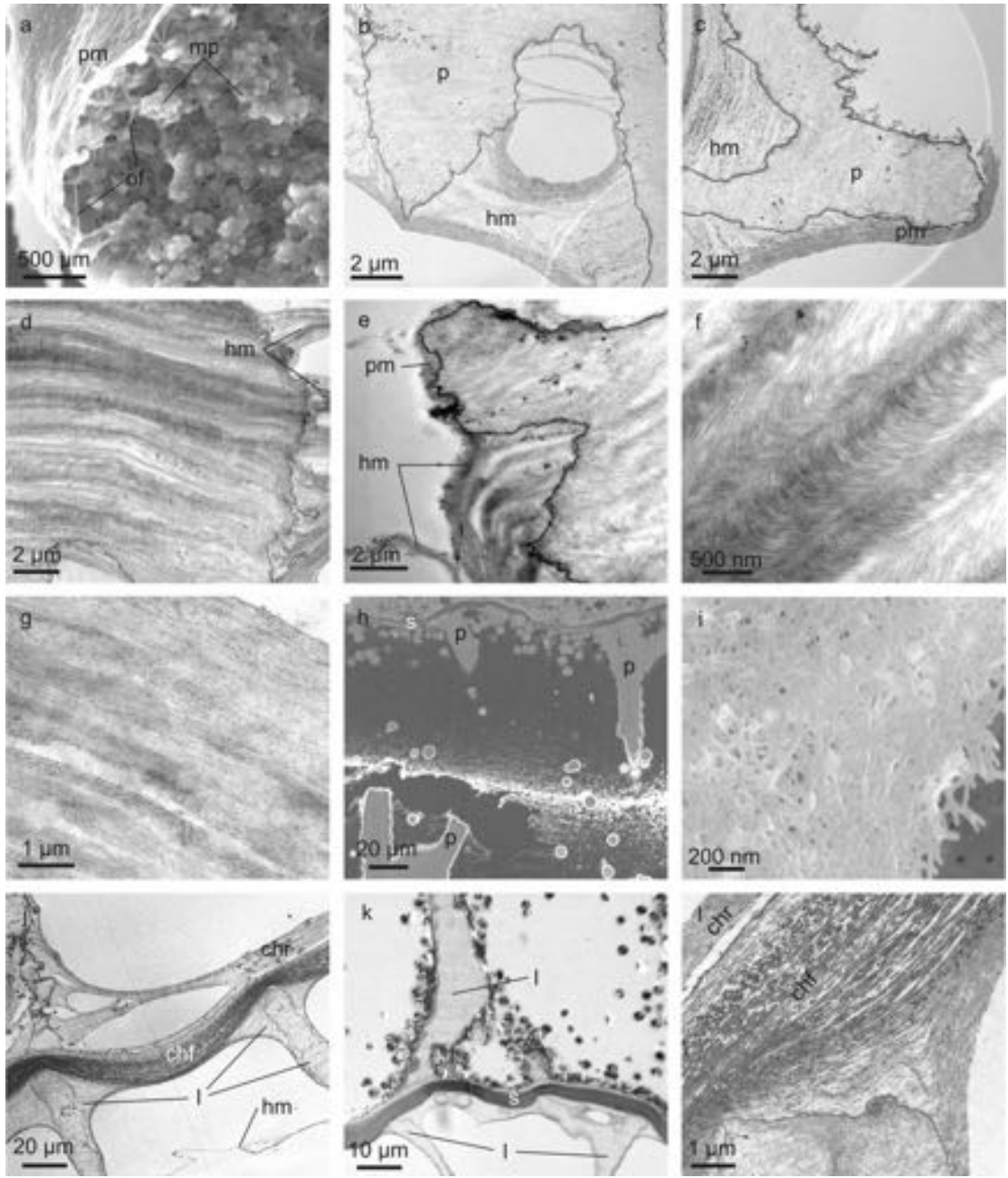

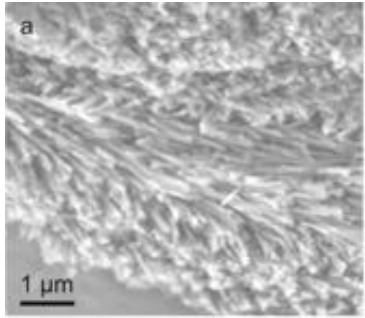
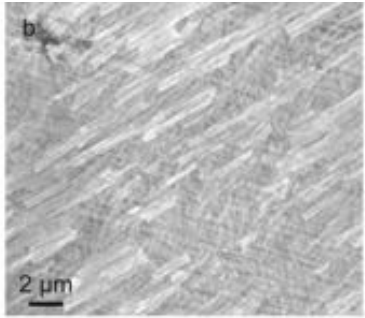
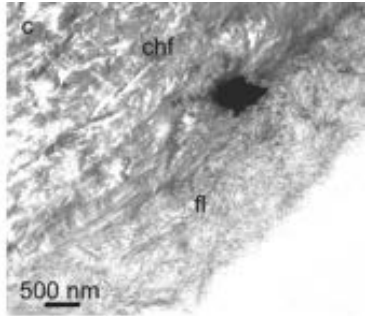
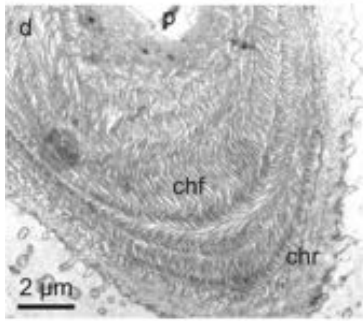
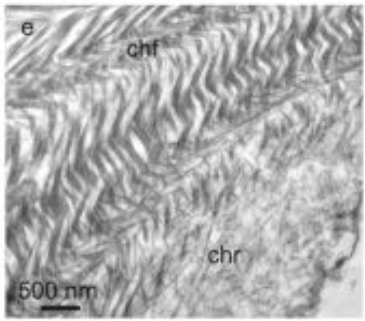
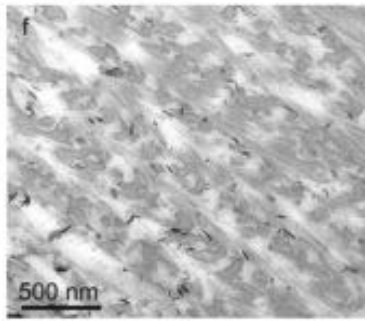
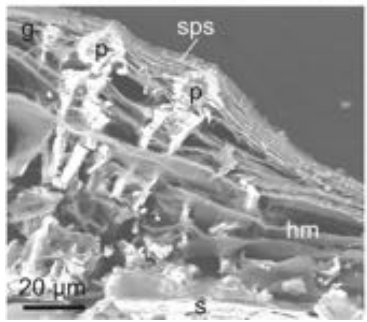
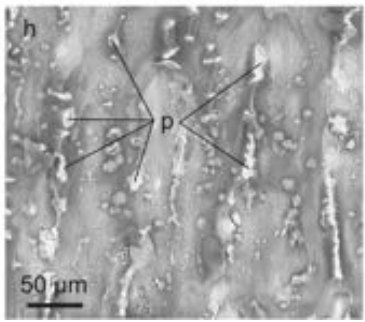
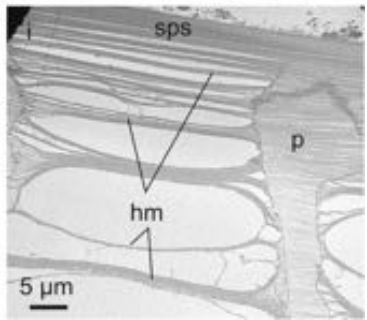

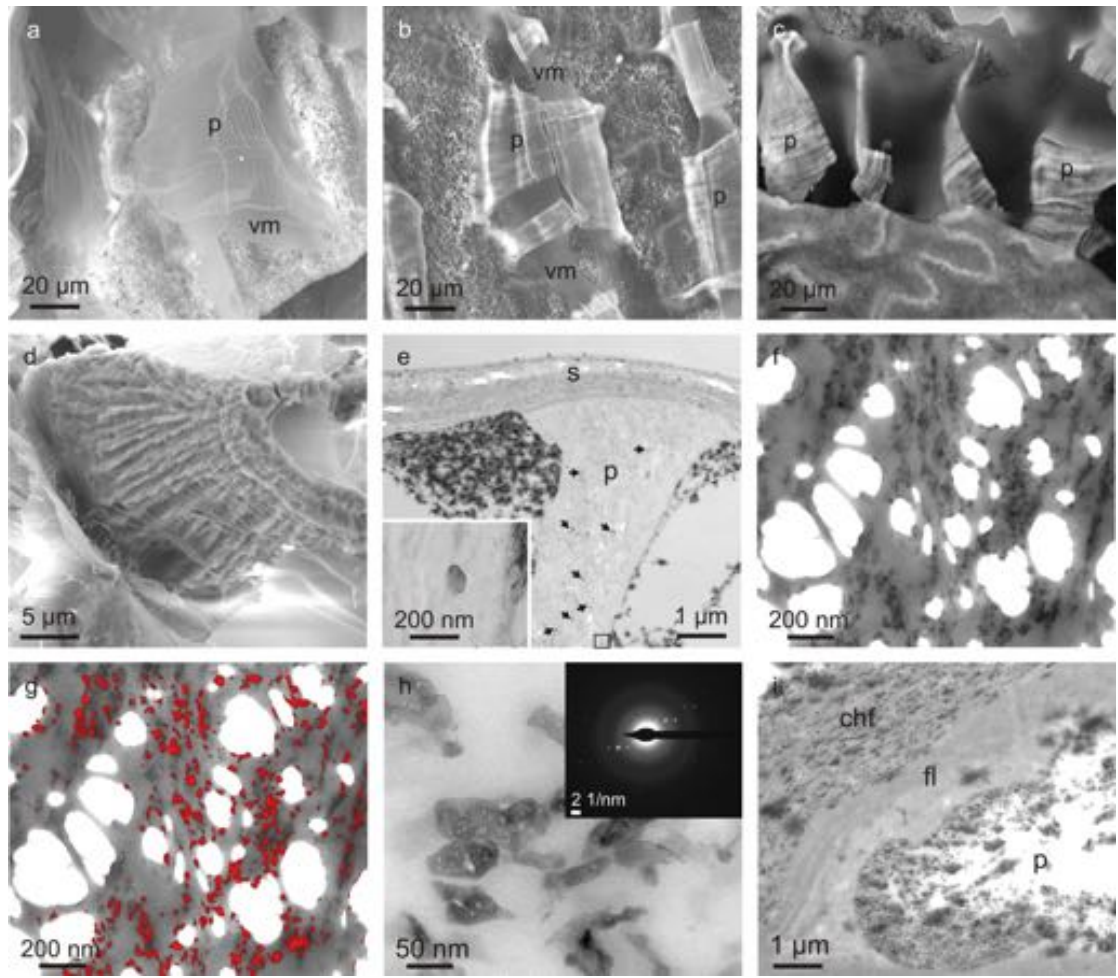

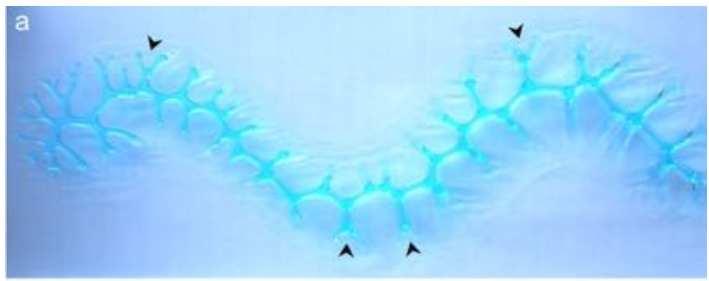

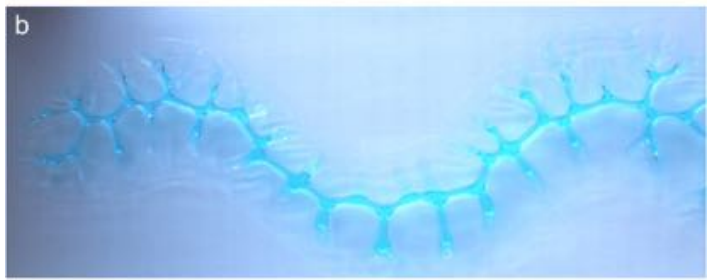

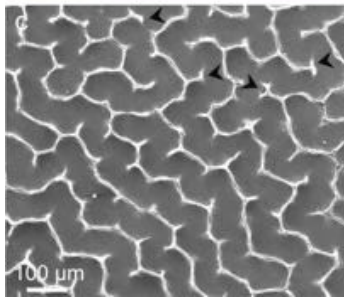
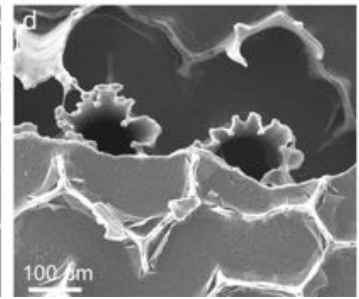

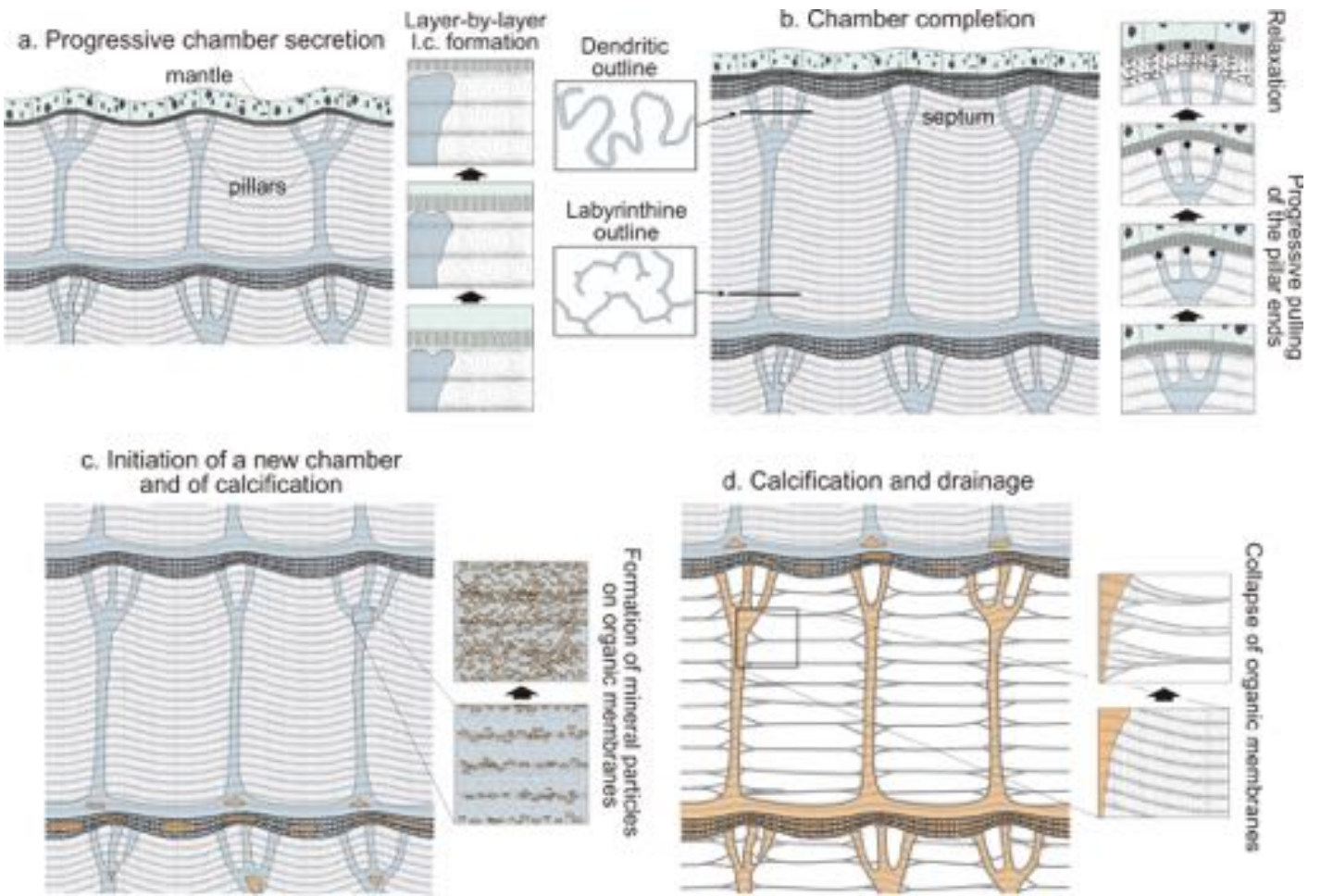

a. Progressive chamber secretion

mantle

pillars

Layer-by-layer l.c. formation

Dendritic outline

Labyrinthine outline

b. Chamber completion

septum

Relaxation

Progressive pulling of the pillar ends

c. Initiation of a new chamber and of calcification

Formation of mineral particles on organic membranes

d. Calcification and drainage

Collapse of organic membranes

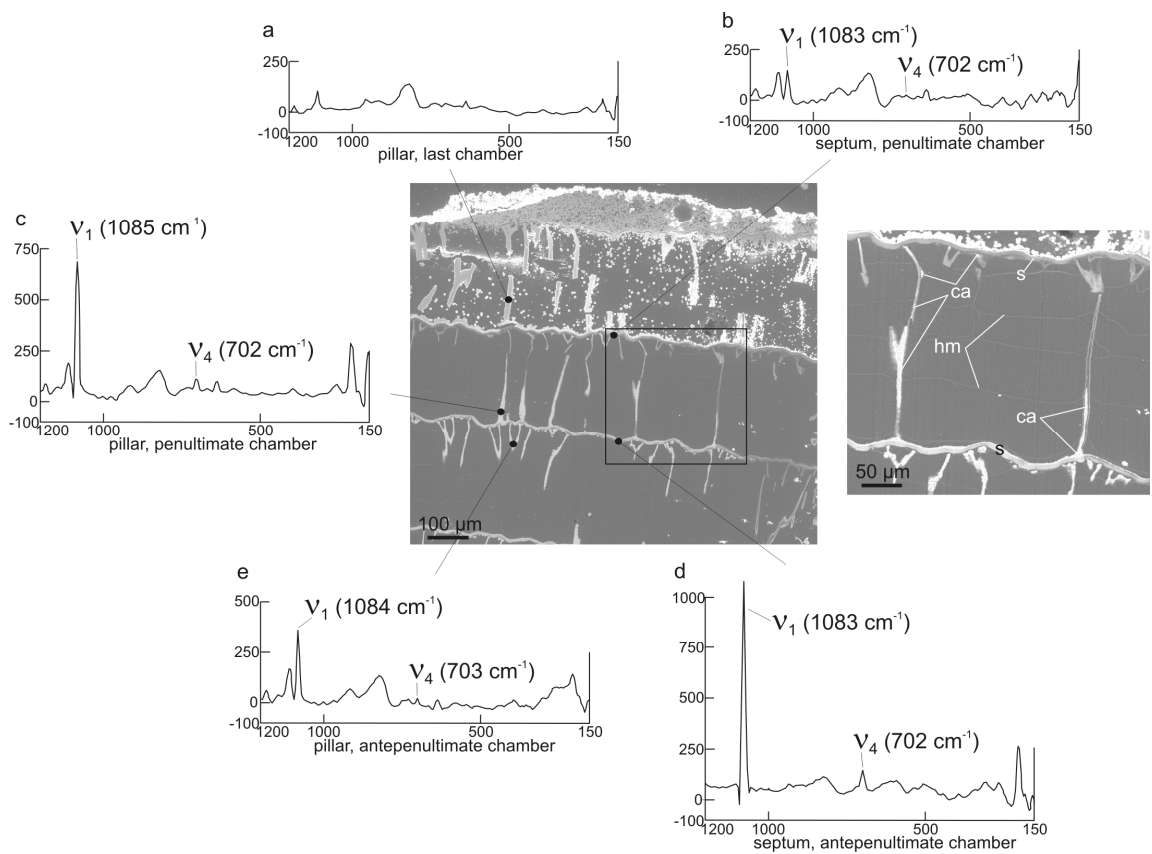

**Figure S1 | Raman analysis of the last formed chambers of an embedded juvenile cuttlebone.** When present, the most intense $\nu_1$ and $\nu_4$ bands of aragonite are indicated. According to the spectra, the last chamber (**a**) is totally unmineralized, the penultimate septum (**b**) is slightly mineralized, and the pillars of the penultimate chamber (**c**) and all the elements of the antepenultimate chamber are mineralized with aragonite (**d**, **e**). The right image is a close up of the framed area, where it can be appreciated that only the bases and some additional spots of the pillars, as well as the penultimate septum, are mineralized. The SEM images have been acquired in back scattered mode. ca= calcified areas, hm= horizontal membranes, s= septa.

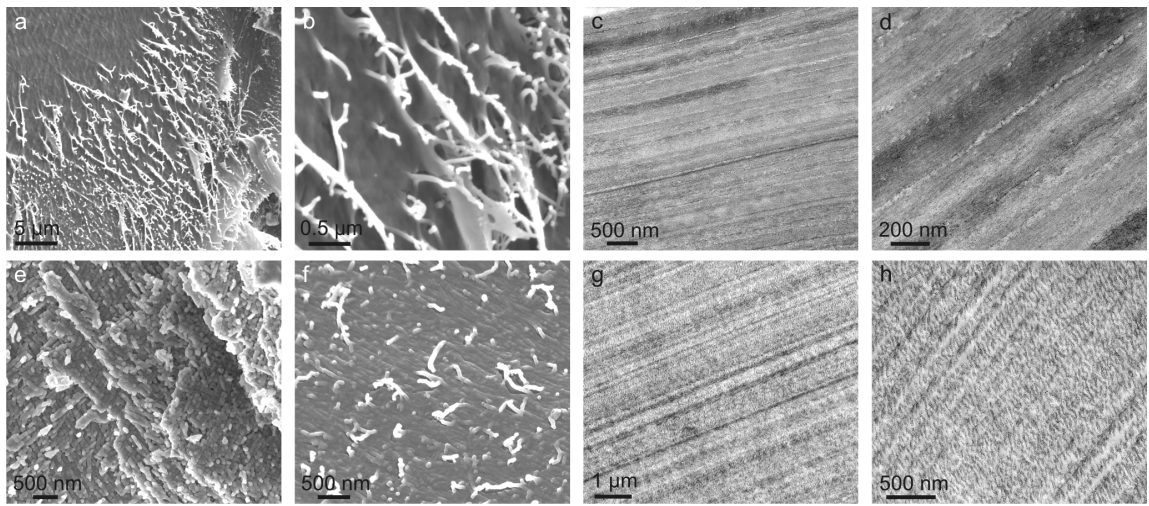

**Figure S2 | Structures of the dorsal shield of the cuttlebone of *Sepia* and of the septum of *Spirula*. a-d**, Dorsal shield of the cuttlebone of *Sepia officinalis*. **a**, SEM view showing the layered aspect. Each layer is made of co-oriented fibres ~50 nm thick. **b**, Close-up view of **a**. **c**, **d**, TEM views; the layered aspect is evident. **e-h**, Septum of *Spirula spirula* (Linnaeus, 1758). **e**, Oblique fracture showing the microstructure composed of layers of co-oriented fibres of aragonite (70-90 nm thick) at high angles to those of adjacent planes. **f**, Surface view of the decalcified septum showing the arrangement of co-oriented organic fibres. **g**, TEM view showing the layered aspect and the arcuate internal distribution of some of the layers. **h**, Detail of **g**, in which the individual fibres of, presumably, chitin, can be discerned; they have thicknesses between 10 and 20 nm.

**Supplementary Video S1 | Tomographic reconstruction of a fragment of the last three chambers of a subadult specimen of *Sepia officinalis*.** The cut out area of the last chamber is the siphuncular area. Only the calcified elements are shown. The video is intended to show (1) the lack of connection of the pillars of the last chamber with the chamber roof, due to incomplete calcification, (2) the peculiar, antler-like morphology of pillars of the siphuncular area and (3) the anteroposterior alignment of the pillars, which is particularly noticeable at their dorsal ends. The venter is to the top.